\documentclass{sf2a-conf2015}
\usepackage{graphicx}
\usepackage{hyperref}
\usepackage[]{natbib}  
\usepackage{epstopdf}

\def\BibTeX{{\rm B\kern-.05em{\sc i\kern-.025em b}\kern-.08em
    T\kern-.1667em\lower.7ex\hbox{E}\kern-.125emX}}
\bibpunct{(}{)}{;}{a}{}{,}  

\begin{document}

\TitreGlobal{proceeding of the SF2A 2015. 7 pages, 2 Fig}


\title{Mapping optically variable quasars towards the Galactic plane}

\runningtitle{Variable QSO}

\author{J. G. Fernandez-Trincado}\address{Institut Utinam, CNRS UMR 6213, Universit\'e de Franche-Comt\'e, OSU THETA Franche-Comt\'e-Bourgogne, Observatoire de Besan\c{c}on, BP 1615, 25010 Besan\c{c}on Cedex, France.}

\author{T. Verdugo}\address{Centro de Investigaciones de Astronom\'ia, AP 264, M\'erida 5101-A, Venezuela.}

\author{C. Reyl\'e$^1$}
\author{A. C Robin$^1$}
\author{J. A. de Diego}\address{Instituto de Astronom\'ia, Universidad Nacional Aut\'onoma de M\'exico, Apdo. Postal 70264, M\'exico D.F., 04510, Mexico.}
\author{V. Motta}\address{Instituto de F\'isica y Astronom\'ia, Universidad de Valpara\'iso, Avda. Gran Breta\~na 1111, Playa Ancha, Valpara\'iso 2360102, Chile.}
\author{L. Vega}\address{Instituto de Astronom\'ia Te\'orica y Experimental (IATE) - C\'ordoba, Argentina.}
\author{J. J. Downes$^{2,}$}\address{Instituto de Astronom\'ia, UNAM, Ensenada, C.P. 22860, Baja California M\'exico.}
\author{C. Mateu$^{2, 6}$}
\author{A. K. Vivas}\address{Cerro Tololo Interamerican Observatory Casilla 603, La Serena, Chile.}
\author{C. Brice\~no$^7$}
\author{C. Abad$^2$}
\author{K. Vieira$^2$}
\author{J. Hern\'andez$^2$}
\author{A. Nu\~nez}\address{Centro de Modelado Cient\'ifico, Universidad del Zulia, Maracaibo 4001, Venezuela.}
\author{E. Gatuzz $^{9,}$}\address{Centro de F\'isica, Instituto Venezolano de Investigaciones Cient\'ificas (IVIC), Caracas 1040, Venezuela.}\address{Escuela de F\'isica, Facultad de Ciencias, Universidad Central de Venezuela, PO Box 20632, Caracas 1020A, Venezuela.}




\setcounter{page}{237}


\maketitle


\begin{abstract}
	We present preliminary results of the CIDA Equatorial Variability Survey (CEVS), looking for quasar (hereafter QSO) candidates near the Galactic plane. The CEVS contains photometric data from extended and adjacent regions of the Milky Way disk ($\sim$ 500 sq. deg.). In this work 2.5 square degrees with moderately high temporal sampling in the CEVS were analyzed. The selection of QSO candidates was based on the study of intrinsic optical photometric variability of 14,719 light curves. We studied samples defined by cuts in the variability index ($V_{index}>66.5$),
	periodicity index ($Q > 2$), and the distribution of these sources in the plane ($A_T,\gamma$), using a slight modification of the first-order of the structure function for the temporal sampling of the survey. Finally, 288 sources were selected as QSO candidates. The results shown in this work are a first attempt to develop a robust method to detect QSO towards the Galactic plane in the era of massive surveys such as VISTA and Gaia.
\end{abstract}



\section{Introduction}

In the past few years the intrinsic variability of the QSO
\citep{rengstorf2004a, rengstorf2004b, rengstorf2006b, schmidt2010, ross2013, graham2014}
has been used as an alternative and efficient selection method to distinguish QSO exhibiting variability from the non-variable stellar locus. Additionally,
it provides unique information about the physics of the unresolved central source. 
This technique is free of bias in comparison with the inherent biases present
in the traditional methods based in specific cuts in the color-color diagram 
\citep{hall1996, croom2001, richards2002, graham2014}.\\

The intrinsic variability of QSO has been observed in different photometric
bands from UV, optical to X-ray. This intrinsic property 
has been proposed as an efficient method of selection 
\citep{rengstorf2006b, macLeod2012, graham2014}, and is commonly quantified using the 
structure fuction such that the amplitude of variability  
changes with time \citep{hughes1992, collier2001, bauer2009, welsh2011}. The 
properties of the variability in QSO have been studied in 
the literature and they depend on the physical properties
of the source, as the presence of radio emission, timescale and others \citep{macLeod2012} and 
the large luminosity of these sources is provided by mass 
accretion onto super massive black holes in its center \citep{Salpeter+1964, Lynden-Bell+1969, Rees+1984}.\\

Both methods are now used to select QSO candidates, especially in surveys with extended time coverage like the 
SDSS Strip 82 \citep{macLeod2012}, MACHO \citep{pichara2012}, Catalina Real-Time Transient Survey \citep{drake2009, graham2014} and 
in the future with Gaia \citep{mignard2012}. The variability selection is a technique with a high degree 
of confidence \citep{graham2014}, and recent studies have shown that variability as method for QSO selection 
is more accurate and has a higher degree of purity in comparison with the use of color-only cuts \citep{morganson2014}.\\

Precise identification of QSOs along the Galactic plane is very valuable for astronomical reference frame purposes.
The large extinction present in the area has meant that all QSO-oriented
surveys have systematically avoided this region, and the scarcity of
confirmed QSO there is evident. Kinematical studies of the Galactic Disk
and Bulge, especially when working in small fields of view,
would benefit enormously by having a dense and deep ``network''
on confirmed QSO, to which tie in their observations.
For large-scale surveys like Gaia, it is also important to have a large enough
number of QSOs identified everywhere on the sky, to improve
the overall quality and spatial uniformity of the QSO reference system.
Finally, the ICRF always will benefit from adding more new QSO,
as the densification of their sources obviously improves
the final accuracy of such fundamental reference frame.\\

The extensive variability survey compiled in the CIDA Equatorial Variability Survey (CEVS) since 2001 has been used so far 
to study a variety of topics, including RR Lyrae stars, T Tauri stars and young Brown Dwarfs, etc. \citep[e.g.][]{Vivas+2004, Briceno2005, Downes+2008, mateu2012}. At high galactic latitudes \citet{rengstorf2004b,rengstorf2004a, Rengstorf2009} used the QUEST Variability Survey data, predecessor to the CEVS, to conduct a variability survey of QSOs over $\sim 190$ sq. deg. However, the CEVS has not been applied to extragalactic studies so far, particularly in the search for QSOs.\\

This paper is organized as follow. In \S\ref{Data} we briefly present the data. The methods are described in 
\S\ref{methods: selection}. The expected contamination is discussed in \S\ref{contamination}. In \S\ref{Conclusion} we present a preliminar conclusion of this research.\\

  \section{Data}
  \label{Data}
  
  The CEVS provides optical multi-epoch information in the $V$, $R$ and $I_c$ photometric bands. For our work we have combined data from the QUEST\footnote{Quasar Equatorial Survey Team} high-galactic latitude survey data, obtained during 1998 to 2001, with the CEVS data collected from 2001 to 2008. The full catalog contains more than $6.5\times{}10^6$ sources, observed multiple times from 1998 to 2008. All observations were obtained with the QUEST mosaic camera (16 CCDs) installed at the 1.0/1.5m J\"urgen Stock Schmidt telescope
  located at the National Astronomical Observatory of Venezuela. The survey has been scanned 476 deg$^2$ of the sky during ten years in a region defined between
  $60\textordmasculine \leq \alpha \leq 140\textordmasculine$ and $-6\textordmasculine \leq \delta \leq 6\textordmasculine$ 
  around the Galactic plane. A detailed description of the survey is given in \citet{mateu2012}.\\ 
  
  We selected one specific section of the catalog (with a large number of observations in each band, $N>10$), restricted to the range 
  $85\textordmasculine \leq \alpha \leq 87.5\textordmasculine$ 
  and $-1.5\textordmasculine \leq \delta \leq -0.5\textordmasculine$, with 14,719 sources covering an area 
  of 2.5 deg$^2$ near the Galactic plane that also has with observations over 1.96 deg$^2$ from the SDSS DR9 \citep{Ahn+2012}. This region in the CEVS has typically about
  30 observations per source.\\

  \section{Selection criteria}
  \label{methods: selection}
  
  In subsections \ref{Method1}, \ref{Method2} and \ref{Method3} below, we describe 
  three methods (modified in this work), proposed in the literature to explore the intrinsic optical 
  variability, periodicity, and the first-order of the structure function, with the main goal to
  separate variable QSO and point sources of non-variability sources (more likely associated with the stellar locus). \\ 
  
  \subsection{Variability}
  \label{Method1}
  
  We adopted a formulation to that of eq. 1 in \citet[][]{rengstorf2006b}, in order to 
  characterize the variability over three optical photometric bands, $V$, $R$ and $I_c$ of the 
  CEVS. For this purpose, we defined the index of variability $V_{index}$ 
  according to the following criteria: (\emph{i}) A minimum of 10 observations was imposed in each photometric band; 
  (\emph{ii}) The index $P(\chi^2)$ given by the CEVS, represents the probability of variability for each photometric band, and its value is related to 
  the $\chi^2$ \citep[e.g.,][]{Vivas+2004}. The $V_{index}$ of a star is redefined in this work as:
  
  \begin{equation}
  V_{index}\equiv \sum^3_{j=1} \frac{\left(N^{j}/N^j_{T}\right)\times \left(1-P(\chi^2_{j})\right) \times{} 100}
  {\sum^3_{j=1}\left(N^{j}/N^j_{T}\right)} = V^{j=1} + V^{j=2} + V^{j=3}, 
  \label{variability1}
  \end{equation}

  \noindent where $j$ indexes over filters ($j=1$ for $V$, $j=2$ for $R$, $j=3$ for $I_{c}$); $N^j$ 
  is the number of observations for the \emph{i-th} source and $N^j_{T}$ is the maximum number of observations
  inside a cone search of 30 arcsec radius centered on the \emph{i-th} source of the catalog. We have computed $V_{index}$, for 14,719 sources, and 
  we have found for each photometric band the following percentages of sources for which
  $V^j\neq0$: $V^{j=1}=30.90\%$, $V^{j=2}=54.77\%$ and $V^{j=3}=65.35\%$. Figure \ref{V_{index}} shows the cumulative probability distribution ($F > V_{index}$) for these 
  14,719 sources (black line) and 39 spectrally confirmed QSO (red line) from the SDSS DR9. The black
  vertical dashed line, correspond to the limit $V_{index} = 66.5$ imposed in this work to separate non-variable sources of 
  variable sources, which is the same value 
  proposed by \citet{rengstorf2006b}. Finally, we selected 1,931 variable sources with
  $V_{index}\geq66.5$, for which the cumulative probability distribution of $F(>66.5)$ is about 13\% and $F(>66.5)=5\%$ for the 
  QSO reported in this region of the sky. \\
  
  \begin{figure}[ht!]
  	\begin{center}
  		\includegraphics[width=0.33\textwidth]{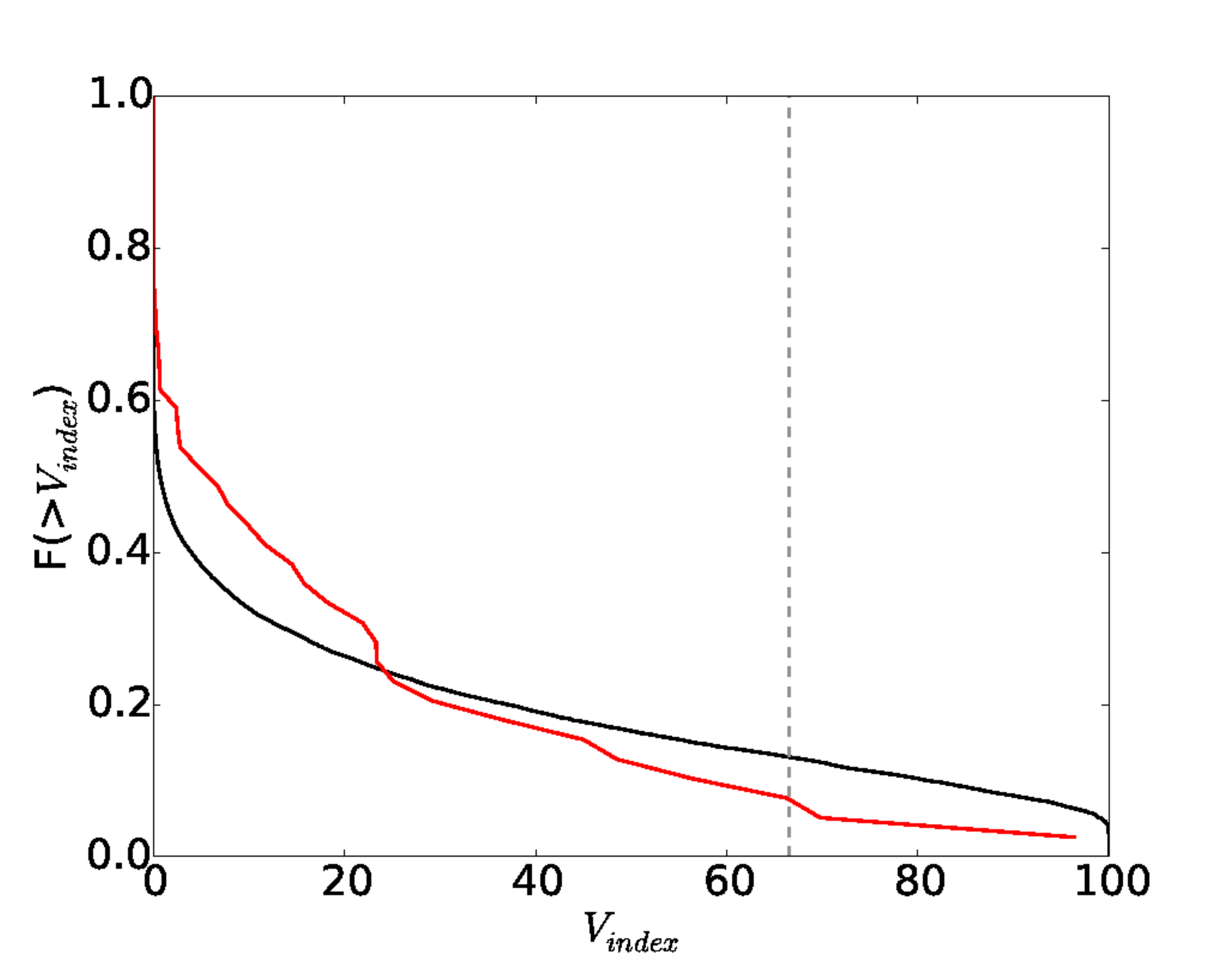}\includegraphics[width=0.33\textwidth]{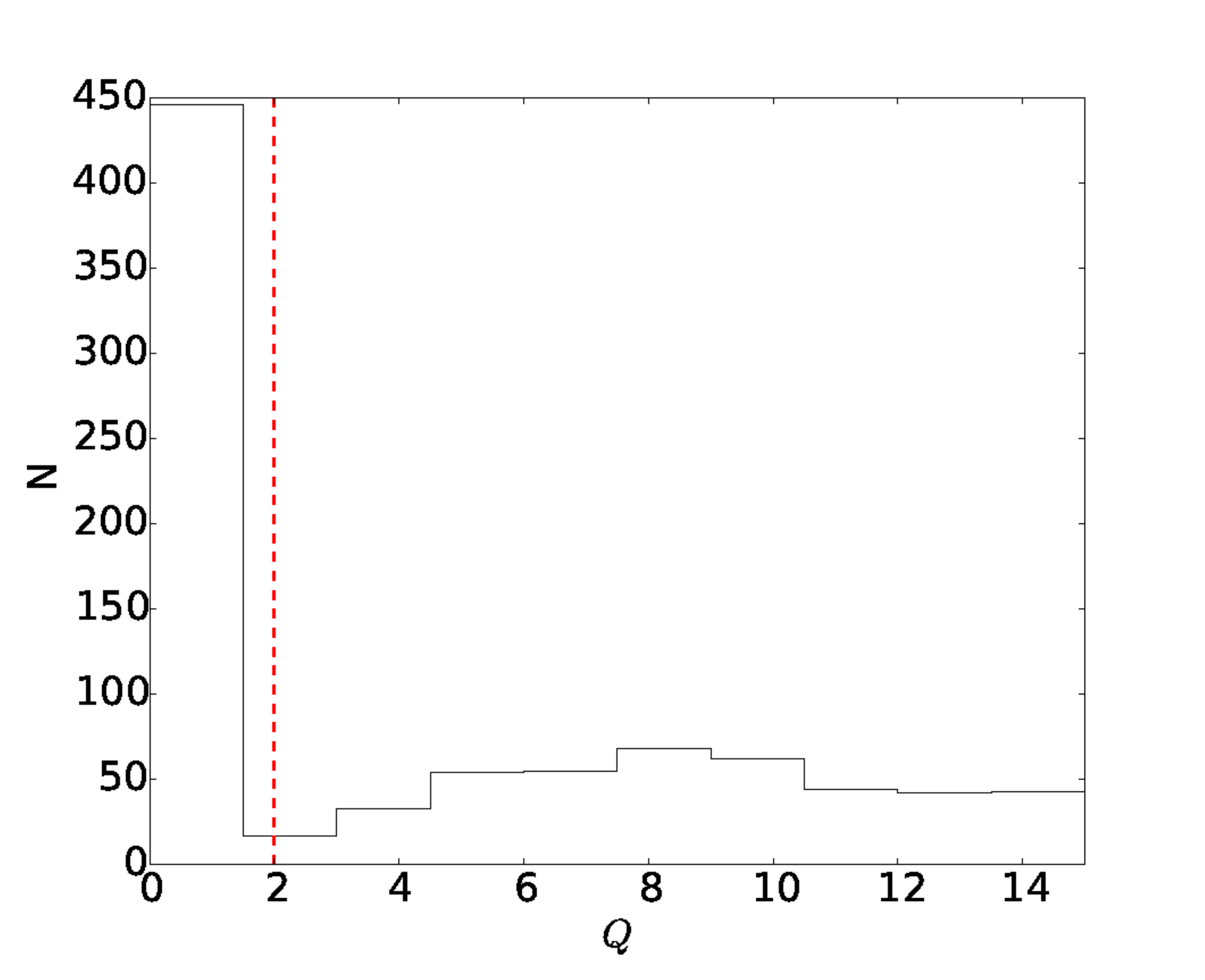}
  		\includegraphics[width=0.33\textwidth]{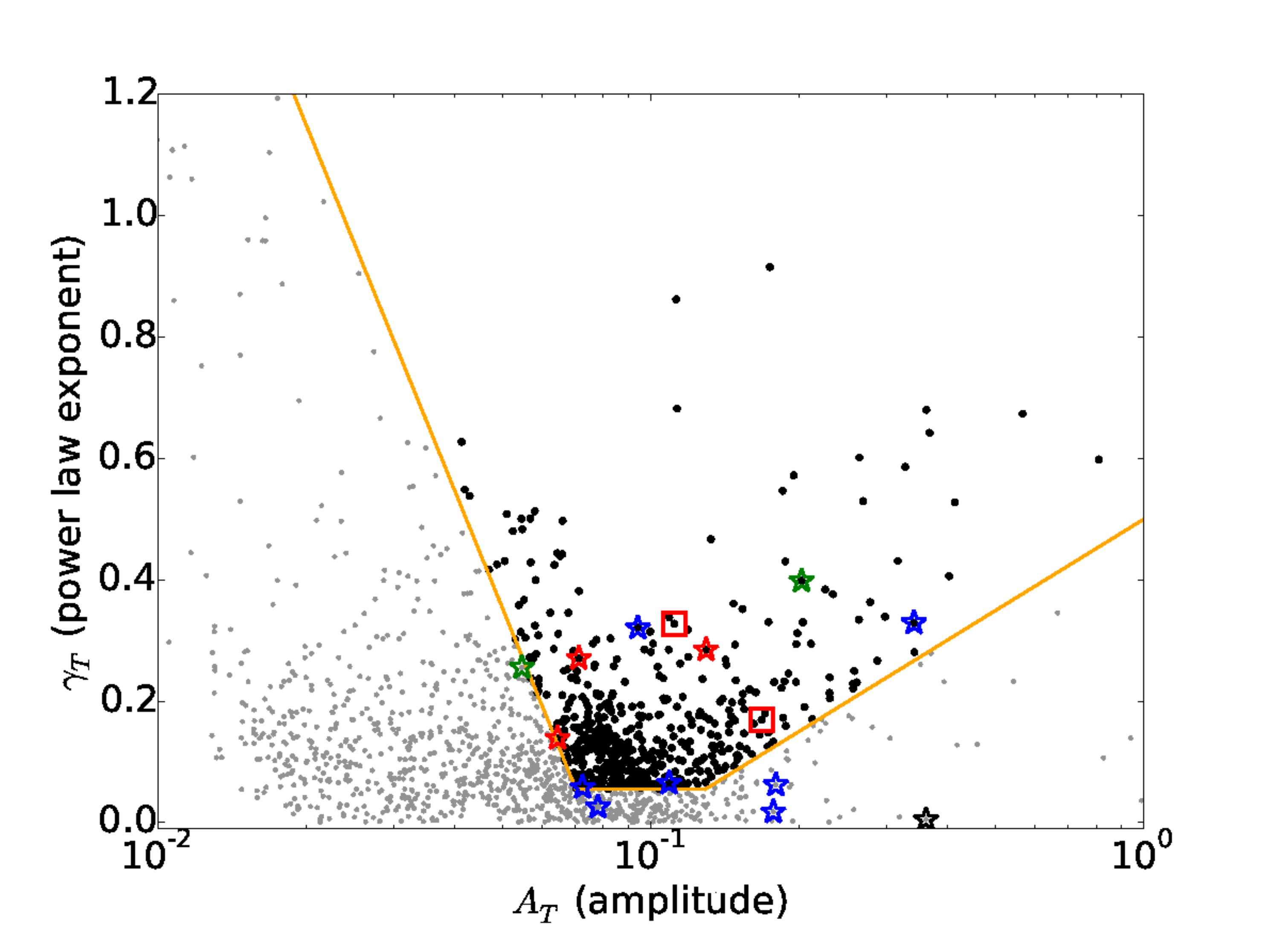}\includegraphics[width=0.4\textwidth]{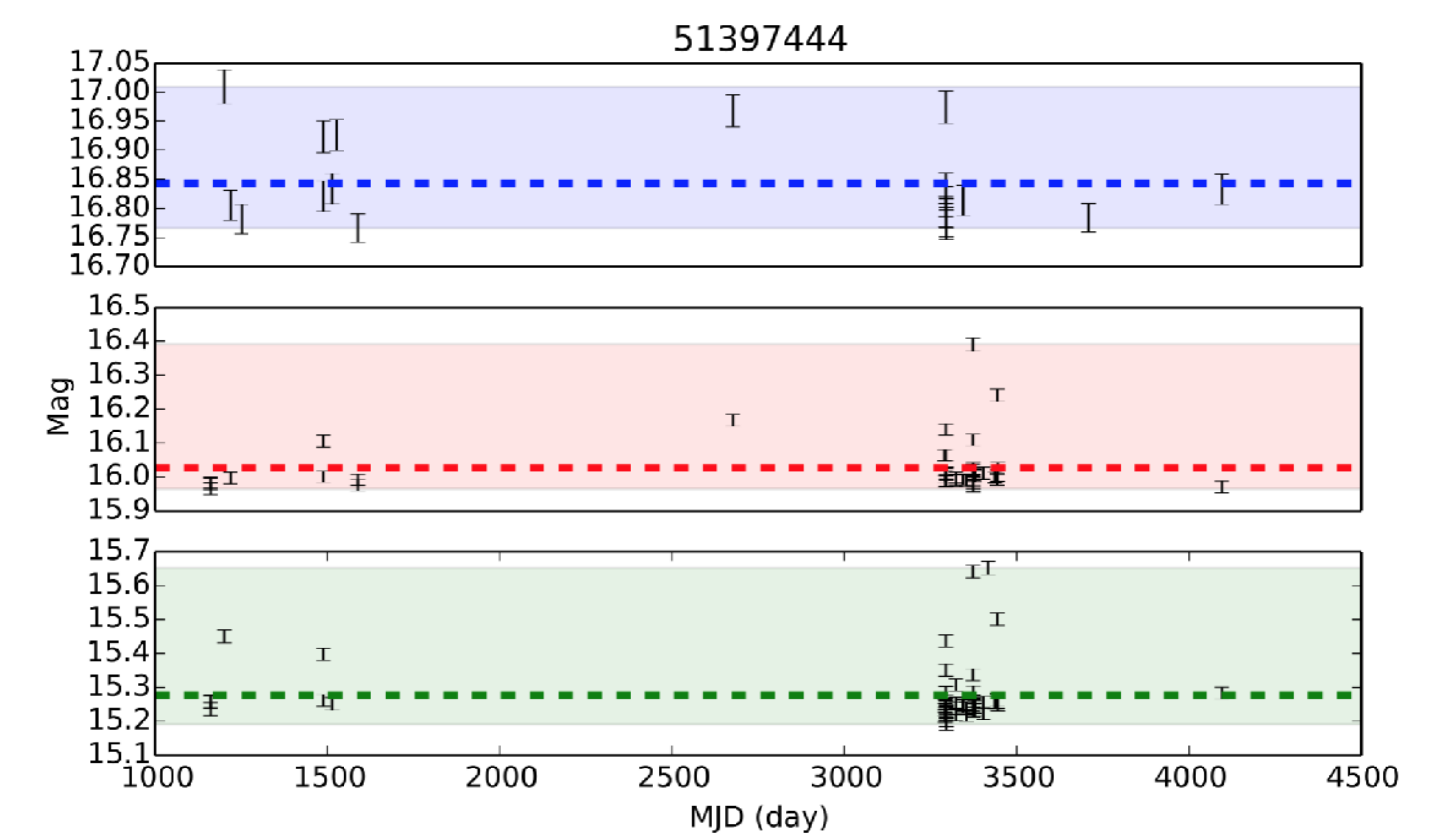}
  	\end{center}
  	\caption{{\bf Top left panel:} Cumulative probability distribution ($F > V_{index}$) for 14,719 sources from the CEVS (black line), and QSO spectrally confirmed in the SDSS DR9 (red line) and observed by CEVS. The vertical dashed 
  		line corresponds to the limit used in this work ($V_{index}=66.5$), to rejected non-variable sources ($V_{index}<66.5$). {\bf Top right panel:} $Q_{i}$ distributions for 1,931 variable sources. QSOs
  		candidates have $Q>2$, meaning that they are both variable and aperiodic sources. {\bf Bottom left panel:} QSO candidates parametrized by
  		the structure function in the plane $(A_{T}, \gamma_{T})$ are
  		shown with black dots and QSO spectrally confirmed are shown in red open squares. 
  		All dots correspond to the sample of 1,481 sources analyzed in the third step (see \S~\ref{Method3}). Contaminants in our sample reported in the literature are:
  		binary stars (blue open stars), RR Lyrae (black open star), 
  		brown dwarfs (green open stars), variable stars in the CATALINA survey (open red stars). {\bf Bottom left panel:} Example of a QSO candidate light-curve. From top to bottom 
  		observations in the photometric bands
  		$V$, $R$ and $I_c$ respectively, are shown.} 
  	\label{V_{index}}
  \end{figure}
  
  \subsection{Periodicity}
  \label{Method2}
  
  For our pre-selection of 1,931 variable sources in \S \ref{Method1}, we redefined the parameter $Q$, in 
  order to separate periodic from aperiodic variable sources. We analyzed each light-curve
  independently, considering a minimum of
  ten available epochs taken from 1998 to June 2008, 
  for the temporal sampling in the light-curves. The parameter $Q$, was redefined in a very similar way as eq. 7 in \citep[][]{rengstorf2006b}:\\
  
  \begin{equation}
  Q\equiv\frac{\sum_{j}\left(N^j/N^j_T\right)\times \left(\frac{\sum_k \left(N^j/N^j_k\right)\times\left(\sigma^j_{T}/\sigma^j_k\right)^2}{\sum_{k}\left(N^j/N^j_k\right)} \right) \times{}{\Delta \langle m^j \rangle_{\rm{max}}}}{\sum_j\left(N^j/N^j_T\right)\times \sigma^j_T}, \\          
  \end{equation}\\
  
  \noindent where the index $j$, $N^j$ and $ N^j_T$ are defined as in the previous subsection; the index $k$ correspond 
  to the number of 15-days intervals within the time series of observations of the star; $N_k$ and $\sigma_{k}$ correspond to the number of observations 
  in each bin and the standard deviation of the magnitudes by bin in the light curve, respectively; 
  $\sigma_T$ correspond to the total standard deviation of the magnitudes. Figure \ref{V_{index}} shows the $Q$ distribution for our pre-selected 1,931 variable sources. We found 
  1,481 sources, and 2 spectrally confirmed QSO from SDSS DR9, satisfying the condition of aperiodicity, that is $Q>2$,
  based in previous studies \citep{rengstorf2006b}, set to minimize contamination by likely periodic variables.\\
  
  \subsection{First-order of the structure function}
  \label{Method3}
  
  So far, we have done a pre-selection of 1,481 sources identified as potential variable and aperiodic QSO in the CEVS. 
  Our final step, consist in identifying the location of QSO candidates in the plane ($A_T,\gamma{}_T$).  
  The first-order of the structure function, which quantify the variability amplitude
  as a function of time, has been used used for this purpose. We computed an equivalent formulation, rewriting Eq. 3 in
  \citet[][]{schmidt2010}: \\
  
  \begin{equation}
  M^{}_{i,l}(\Delta{}t_{i,l})\equiv \left\langle \sqrt{\frac{\pi}{2}} |\Delta{}m_{i,l}| - \sqrt[]{\sigma_{i}^2+\sigma_{l}^2} \right\rangle _{\Delta{}t},
  \end{equation}
  
  \noindent where $\Delta{}m_{i,l}$ is the measured magnitude difference between observation $i$ and $l$, and
  $\sigma_{i}$ and $\sigma_{l}$ are the photometric errors, being $\Delta{}t_{i,l}$ the time difference between 
  two observations. Thus, the average is taken over all epoch pairs $i$, $l$ that falls in the bin $\Delta t$. In the same way as \citet{schmidt2010}, we parametrized the structure function as:\\
  
  \begin{equation}\label{eq:SF}
  M^{mod}_{i,l}(\Delta{t_{i,l}} \mid A^j, \gamma{}^{j})\equiv A^{j} \left( \frac{\Delta{t_{i,l}}}{1 \rm{yr}} \right)^{\gamma{}^{j}},
  \end{equation}
  
  \noindent where the subindex $j$ stands for filter, and can take the values, V, R, I, and:

  \begin{equation}\label{eq:AT}
  A_{T}=\frac{\sum_{j}\left(N^j/N^j_T\right)\times A_{j}}{\sum_{j}\left(N^j/N^j_T\right)} ,         
  \end{equation}

  \noindent and
  
  \begin{equation}\label{eq:Gamma}
  \gamma_{T}=\frac{\sum_{j}\left(N^j/N^j_{T}\right)\times \gamma_{j}}{\sum_j\left(N^j/N^j_T\right)^{j}},          
  \end{equation}
  
  \noindent Figure \ref{V_{index}} show the distribution $A_T$ and $\gamma{}_T$ for the pre-selected 1,481 sources, and the best fitting values of Eq.~(\ref{eq:SF}) to Eq.~(\ref{eq:Gamma}) to the data. 
  We defined a QSO selection box as Eq. 7 to Eq. 9 in \citet{schmidt2010}, which led us to select a sample of 288 QSO candidates.\\
  
  The final sample of candidates is shown in the Table \ref{TABLEQSO},
  containing all the relevant information for each star in it: column 1-10 list ID (CEVS-QSO-XXX; notation adopted in this work), mean $V$, $R$ and $I_c$ magnitudes, number 
  of times each star was observed in each photometric band and amplitud. The index of variability and aperiodicity are presented in columns 11 and 12 respectively, and 
  column 13 and 14, correspond to the parameters of the structure function ($A_T$ and $\gamma_{}$, respectively). Table \ref{TABLEQSO} is published in its entirety in a public repository\footnote{\url{http://fernandez-trincado.github.io/Fernandez-Trincado/simulations.html}}\footnote{jfernandez@obs-besancon.fr}.
  A sample of the table is shown here for guidance regarding its content.
  
  \begin{table*}[ht!]
  	\setlength{\tabcolsep}{1.7mm}  
  	\centering
  	\caption{Photometric parameters of QSO candidates from the CIDA Equatorial Variability Survey (CEVS). }
  	\label{TABLEQSO}
  	\begin{tabular}{@{}ccccccccccrrcc@{}}
  		\hline
  		\hline
  		ID       & $\langle V \rangle$  & $\langle R \rangle$    & $\langle I \rangle$    & Nv  &  Nr  &  Ni   &  AmpV   &   AmpR    &   AmpI   &   Vindex   &   Q        &   $A_T$        &   $\gamma_T$ \\         
  		&        [mag]         &         [mag]          &        [mag]           &     &      &       &  [mag]  &   [mag]   &   [mag]  &            &            &                &               \\
  		\hline
  		\hline    
  		CEVS-QSO-001    & 19.323               & 18.045                 & 17.855                 & 11  & 41   & 79    & 1.009   & 1.019     & 0.951    & 100.00     &   58.671   & 0.109          & 0.285 \\
  		CEVS-QSO-002   & 19.449               & 19.010                 & 18.299                 & 18  & 43   & 70    & 0.566   & 0.412     & 0.696    &  99.99     &   59.737   & 0.068          & 0.253 \\
  		CEVS-QSO-003    & 19.371               & 18.791                 & 18.037                 & 16  & 47   & 91    & 0.875   & 0.670     & 0.897    & 100.00     &  115.831   & 0.088          & 0.232 \\
  		CEVS-QSO-004    & 19.438               & 18.699                 & 17.644                 & 9   & 45   & 87    & 1.072   & 0.584     & 0.591    & 100.00     &  124.065   & 0.066          & 0.126 \\
  		CEVS-QSO-005     & 18.747               & 18.450                 & 17.170                 & 17  & 49   & 99    & 0.947   & 0.493     & 0.319    &  73.36     & 1243.269   & 0.049          & 0.425 \\
  		CEVS-QSO-006     & 19.182               & 17.886                 & 16.372                 & 26  & 52   & 90    & 0.455   & 0.454     & 0.357    & 100.00     &  986.204   & 0.341          & 0.329 \\
  		CEVS-QSO-007     & 19.505               & 18.916                 & 18.172                 & 3   & 32   & 36    & 0.431   & 0.508     & 0.517    &  99.72     &   28.757   & 0.115          & 0.071 \\
  		CEVS-QSO-008   & 18.999               & 18.545                 & 17.540                 & 30  & 50   & 78    & 0.703   & 1.012     & 0.607    &  91.33     &   54.579   & 0.062          & 0.346 \\
  		CEVS-QSO-009   & 19.632               & 18.743                 & 17.851                 & 14  & 43   & 69    & 0.563   & 0.569     & 0.561    &  99.99     &    6.741   & 0.132          & 0.467 \\
  		\hline 
  		\hline
  	\end{tabular}
 \end{table*}

  \section{Expected Milky Way contamination}
  \label{contamination}
  
\begin{figure}[ht!]
  	\begin{center}
  		\includegraphics[width=0.5\textwidth]{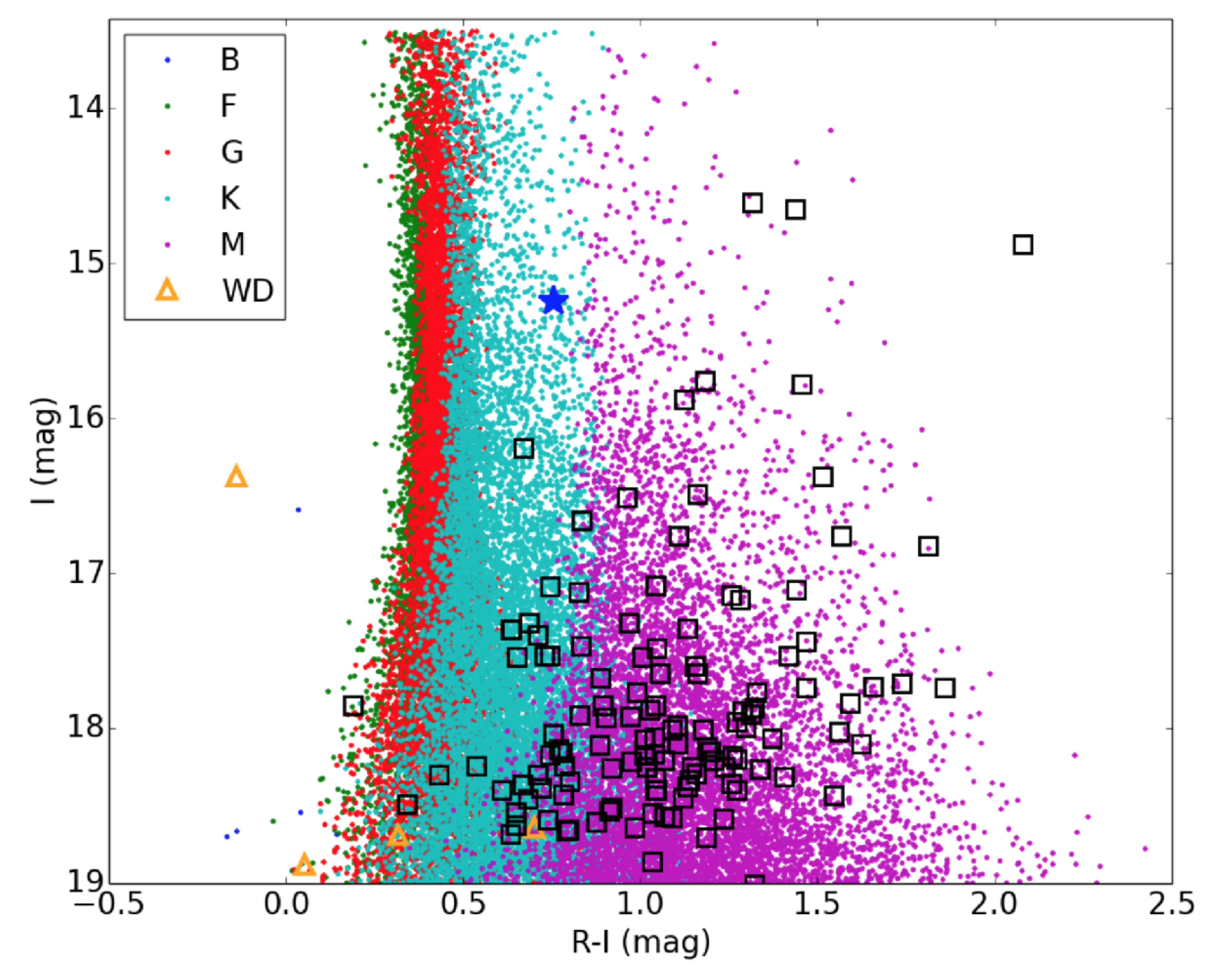}
  	\end{center}
  	\caption{Color magnitude diagram for QSO candidates (black symbols) with I and R information.
  		The blue star refers to object ID 51397444 showed in Figure \ref{V_{index}}. The simulation of the Besan\c{c}on galaxy model is presented 
  		with dots of different colors for each stellar population, which are labelled by spectral type (B, F, G, K, M and white dwarfs WD).} 
  	\label{BGM}
  \end{figure}

  We have compared the synthetic colour-magnitude diagram computed with the Besan\c{c}on galaxy model (hereafter BGM) \citep{Robin2003, Robin2014},
  for the same line-of-sight and solid angle studied in this paper. The 
  simulation was generated taking the selection function of the data into account. Since we are near the Galactic plane, we expect a high degree of contamination (foreground stars), 
  not present in previous surveys of QSO using variability. The absence of a complete catalog of confirmed QSOs in this part of the Galaxy makes it difficult to 
  estimate such contamination, a key point to validate our methodology and its application in future surveys like Gaia or VISTA. At this
  moment we are in the process of determining this contamination, since the CEVS is a not homogeneous survey. However we can do a simple and rough estimate of the more likely contaminants in our sample, namely K and M-type stars, using the BGM and comparing the synthetic colors with those observed in our sample candidates (see Figure \ref{BGM}).

  \section{Conclusion}
  \label{Conclusion}
  
  In this work, we have used the multiepoch data in the large scale CEVS to search for low-galactic latitude QSOs by their intrinsic optical variability, using an alternative formulation 
  similar to the proposed in the literature, in order to detect QSO candidates from 
  inhomogeneous temporal sampling.\\
  
  We have selected 288 QSO candidates according to their variability, aperiodicity and parameters 
  of the structure function, over an area of 2.5 deg$^2$. Extrapolating these results 
  to the full CEVS with an total area of $\sim500$ deg$^2$, we estimate that is possible to detect 52,000 new QSO candidates in this survey. However, we have shown that the methods presented in this work 
  are sensitive to variations in the temporal sampling of the light curves. This 
  explains the fact our method rejects 95\% of the QSO spectroscopically confirmed in the region 
  in common with the SDSS DR9 survey. Follow-up spectroscopic observations for our QSO candidates are currently conducted at the  REOSC Spectrograph installed on the 2.15-m telescope at CASLEO, Argentina. \\
  
  In the near future, large spectroscopic surveys as Gaia may help to confirm QSO sources, selected from variability surveys towards the Galactic plane, allowing us to quantify the selection efficiency.\\
  
  \begin{acknowledgements}
  	J.G.F-T is currently supported by Centre National d'Etudes Spatiales (CNES) through Ph.D grant 0101973 and the R\'egion de Franche-Comt\'e, and by the French Programme National de Cosmologie et Galaxies (PNCG). 
  	This research was supported by the Munich Institute for Astro- and Particle Physics (MIAPP) of the DFG cluster of excellence "Origin and Structure of the Universe".
  	V.M. acknowledges the support from FONDECYT 1120741, and Centro de Astrofísica de Valparaiso. 
  \end{acknowledgements}

\bibliographystyle{aa}  
\bibliography{Fernandez1.tex} 

\end{document}